\documentclass[a4paper, 12pt]{article}

\usepackage[latin1]{inputenc}
\usepackage{graphicx, amssymb, amsfonts, amsmath, mathrsfs}
\usepackage[english]{babel}
\usepackage[official]{eurosym}
\usepackage{caption}[2008/04/01]
\usepackage{newcent}
\usepackage{rotating}
\usepackage{lineno}
\usepackage{setspace}
\usepackage{tabularx}
\usepackage[flushmargin]{footmisc} 
\usepackage{color}
\usepackage{lscape,array,hhline,pstricks,fancyhdr}
\usepackage{lineno}
\usepackage[hscale=1,vscale=1]{geometry}
\usepackage{geometry}
\usepackage{natbib}
\usepackage{url}
\usepackage{threeparttable}
\usepackage{graphicx}
\usepackage{subfigure}

\title{The demand for road transport in China: theoretical regularity and model selection}

\author{Li LIU$^{1}$ and Ling-Yun HE$^{2,3,4,}$ \footnote{Dr. HE is a full professor of applied economics and China studies. LIU is a Ph.D. candidate supervised by Dr. HE.  The authors contribute equally in the project. HE conceived the whole project and designed the scenarios. LIU calculated the results under Dr. HE's supervision. HE and LIU co-wrote the manuscript.  The authors would like to thank Prof. SERLETIS Apostolos from the University of Calgary, Canada, Prof. CHANG Dongfeng from Shandong University, and all other colleagues from China Agricultural University, JiNan University, and the University of Calgary, for all their warm helps, constructive suggestions and pertinent comments. This project is supported by the National Natural Science Foundation of China (Grant Nos. 71273261 and 71573258), and China National Social Science Foundation (No. 15ZDA054).}       \\ \small 1. College of Economics and Management, \\ \small China Agricultural University, Beijing 100083, China\\ \small 2. School of Economics, JiNan University, Guangzhou 510632, China\\ \small 3. Institute of Resource, Environment and Sustainable Development Research, \\ \small JiNan University, Guangzhou 510632, China \\ \small 4. School of Economics and Management, \\ \small Nanjing University of Information Science and Technology, Nanjing 210044, China\\\small * Corresponding author. \\ \small Email: lyhe@amss.ac.cn}

\date{}

\begin{document}

\maketitle

\begin{abstract}
China's road transport sector is found to be one of the major emitters, and responsible for serious air pollution and huge pubic health losses. One important parameter for determining the consequences of transport demand shocks for the macroeconomy, air pollution and public health is the elasticity of the demand for transport. Most published studies that use flexible functional forms have ignored the theoretical regularity conditions implied by microeconomic theories. Moreover, even a few studies have checked and/or imposed regularity conditions, most of them equate curvature alone with regularity, thus ignoring or minimizing the importance of other regularities. And then, the results appear biased and may in fact be biased. Therefore, we select three of the most widely used flexible functional forms, the Rotterdam model, the Almost Ideal Demand System (AIDS), and the quadratic AIDS (QUAIDS) to investigate the demand for road transport in China using recent annual expenditure data, over a 13 year period from 2002 to 2014, on three expenditure categories in the transportation sector: private transportation, local transportation and intercity transportation. We pay more attention to theoretical regularity and believe that unless regularity is attained by luck, flexible functional forms should always be estimated subject to regularity and all regularity conditions are equally important. Estimation shows that the AIDS model is the only model that is able to provide theoretically consistent estimates of the residents demand for road transport in China. Our estimates show that the private transportation is a luxury among the transportation goods, and is elastic in price changes relatively. The empirical results imply that the private and the local transportation, the local and intercity transportation are gross complements. And, the private transportation is a substitute for the intercity transportation, while the intercity transportation is a complement of the private transportation.
\end{abstract}

\indent {\small \emph{Keywords}: Road transport demand; Theoretical regularity; Selection; Estimation}\\

\newpage
\begin{onehalfspace}

\section{Introduction}\label{section_intro}

As a transitional economies, China's long-term extensive economic growth mode has resulted in serious air pollution. Environmental cost of economic development continue to rise, even more, causing more discussion, such as the ``demographic dividend". Many economists attribute the high-speed growth of Chinese economy since the reform and opening up to the release of the ``demographic dividend". But environmental degradation could accelerate the subsidence of ``demographic dividend". The world health organization (WHO) report the average concentrations of PM2.5 in 2973 cities arranged from high to low. There are 30 cities in China at the top one hundred cities\footnote{ Data source: World Health Organization: WHO's Urban Ambient Air Pollution database-Update 2016. \url{http://www.who.int/phe/health_topics/outdoorair/databases/AAP_database_summary_results_2016_v02.pdf?ua=1 }}. China's road transport sector is found to be one of the major emitters, and responsible for serious air pollution and huge residents' health losses ( He and Chen, 2013; Chen and He, 2014 ), especially in urban areas ( Hao and Wang, 2005; Guo et al., 2010 ).

Since the year of 1978 when Chinese central government began to promote its momentous economic reforms, China has been keeping a miraculously high speed of economic growth. At same time, people's living standards have been significantly improved, with the urbanization level increasing from 17.92 percent in 1978 to 54.77 percent in 2014
. What's more, the China's road transport sector has enormous growth in the road transport infrastructure investment, road mileage and stock of vehicles
. In particular, the last decade has witnessed a dramatic increase of the ownership of private vehicles. All of this cause the rapidly growing road transport demand of the Chinese residents. As one of the main sectors of the economy, road transport has a significant impact on the economic development of the country. The increase in the development of transport and transport infrastructure itself definitely develops the economic potential of the country, strengthens its economic, social cohesion and economic competitiveness. Road transport, as the carrier of  production, distribution, exchange and consumption, is an important factor in the economic development. In addition, it offers great convenience for people's travel.

However, the rise of transport demand brings huge negative externalities and social and economic costs. Automobile exhaust contains a lot of toxic and harmful gas. It is easier for the people living in urban where high concentrations of people live and a huge number of cars drive, to breathe heavy traffic fumes. Automobile exhaust poses a great threat to human health, and even lead to death. In addition, the automobile exhaust can aggravate the greenhouse effect, causing tremendous environmental pressure. Rising traffic demand will drive the fuel demand rising, while China is a country with a shortage of oil, and high external dependency (More than 60\% of the oil needed will rely on foreign supplies). This not only brings the burden to the national economy, but also, to a certain extent, affects the energy security of China.

At present, the problem of transportation-related pollution is increasingly more serious in China. In 2015, the vehicle population reached 279 million, producing 30\% of national nitrogen oxide emissions. The road transport sector has become an important source of air pollution in China and is an important reason causing haze and photochemical smog pollution. Especially, in large cities such as Beijing and Shanghai and the eastern densely populated areas, the contribution of vehicles to fine particulate matter concentration reached about 30\%, in extreme adverse weather conditions, the contribution will even reach more than 50\%\footnote{ Data source: Ministry of Environmental Protection of the People's Republic of China: China Vehicle Emission Control Annual Report, 2016. \url{http://www.mep.gov.cn}}. Also, due to the motor vehicle driving in densely populated areas, emissions directly threaten people's health. One hundred million new motor vehicles are estimated in China for the next five years. As a result, there will be new car gasoline and diesel consumption by 100 million to 150 million tons, which will bring enormous pressure to the atmospheric environment and energy security. Given the significance of decreasing Chinese transportation-related fuel consumption and  harmful gas emissions, it is especially important to investigate the demand for road transport in China


One important parameter for determining the consequences of transport demand shocks for the macroeconomy, air pollution and residents' health is the elasticity of the demand for transport. If the demand for transport is very inelastic (so that when the price or expenditure increases by a certain percentage, the quantity demand decreases by a much smaller percentage), transport price or expenditure changes will have significant effects on the level of economic activity through their effect on consumption. For example, sharp increases in transport price will significantly reduce disposable income, causing consumers to curb expenditures across consumption categories. Moreover, higher transport prices saddle businesses with higher transportation coats, causing them to pass them on to their customers in the form of higher product prices.

Since the middle of 1950s, empirical research has focused on the demand system that is consistent with the requirements of the neoclassical theory. Before the emergence of linear expenditure system model ( Stone, 1954a, b ), the demand analysis mainly focused on the single equation estimation, and, with little attention to the consistency of the basic theory. Stone developed a complete demand system that is consistent with the neoclassical theory, and use the data from the United Kingdom to merge consumer goods into a manageable components, so as to achieve the system model estimation. Due to the consumer preferences additivity hypothesis, namely marginal utility from a commodity is independent to the marginal utility from other commodities, the system model of Stone limits properties of the correlation between commodities. As a result, all the goods are substitutes for each other. This may be too simplistic and idealized. Strotz (1957) expands the idea of complete expenditures and use it for a multi-level system, making the feasibility of the estimation of the demand system model from the theory to the economic metrology. In addition, the demand system model can also test the assumptions of the demand theory. Barten (1969), Byron (1970a, b) and Deaton (1978) all discussed the theoretical test methods in details. During the last thirty years it has become standard to use flexible functional forms for empirical consumer demand analyses, such as the Rotterdam, AIDS, QUAIDS, translog and Minflex Laurent model. Iootty et al.(2009) investigate and explain the performance of the Brazilian demand for automotive fuels in the period 1970-2005 through the estimation of the AIDS model. Based on the Translog cost function, Angulo(2014) estimate the water demand in the hotels and restaurants sector.

\begin{table}[h!]
\setlength{\abovecaptionskip}{0pt}
\setlength{\belowcaptionskip}{10pt}
\centering
\begin{threeparttable}[b]
\scriptsize
\captionsetup{labelfont=bf,textfont=normalfont,justification=raggedright,labelsep=space,labelsep=newline,singlelinecheck=off}
\caption{Summary of flexible functional forms:estimation of transport demand} \label{tab1}
\scriptsize
\begin{tabular}{lcccccccc}
\hline
Author    & Model used & Positivity & Monotonicity  & Curvature &Adding-up  & Homogeneity & Symmetry  \\
\hline
Iootty et al.(2009)   &AIDS &No &No &No &Test  &Test &Impose  \\
Sun and Ouyang(2016)  &AIDS &No &No &No &Impose  &Impose &Impose  \\
Ghalwash(2008)  &QUAIDS &No &No &No &Impose  &Impose &Impose  \\
Petr(2014) &QUAIDS &No &No &Test &Impose  &Impose &Impose  \\
Aepli(2014) &QUAIDS &No &No &Observe &Impose  &Impose &Impose  \\
Fashogbon et al.(2013)&QUAIDS &No &No &No &Impose  &Impose &Impose  \\

\hline

\end{tabular}
\end{threeparttable}
\end{table}

The usefulness of flexible functional forms, However, depends on whether they satisfy the theoretical regularity. In this regard, as Barnett (2002) put it, without satisfaction theoretical regularity conditions, the second-order conditions for optimizing behavior fail, and duality theory fails. The resulting first-order conditions, demand functions, and supply functions become invalid.' However, in the literature there has been a tendency (see Table 1) to ignore or not to report theoretical regularity, which makes the result of parameter and elasticities estimation invalid or less rigorous. According, for example, to Serletis and Shahmoradi (2007), only three out of fourteen studies in the monetary asset demand literature since 1983 have addressed theoretical regularity issues. This neither meet the theoretical regularity conditions of neoclassical theory nor go with the economic reality, which attract the attention of many scholars. Chang and Serletis (2014) investigate the demand for gasoline in Canada using three of the most widely used locally flexible functional forms and impose local curvature to produce inference consistent with neoclassical microeconomic theory. Wolff and Heckelei(2010) extend upon a currently available estimation method for imposing curvature and monotonicity on flexible functional forms. However, they pay more attention to curvature and weaken other theoretical regularity, which is equally important. Therefore, we treat each regularity equally in order to be consistent with the neoclassical theory and the economic reality. And then, we use the data of road transport in China to do a simple example, having a deeper understanding of each regularity.

The rest of the paper is organized as follows. Section 2 provides the theoretical regularity. Section 3 follows the demand systems approach and discusses three flexible functional forms, paying attention to the imposition and test of regularity. Section 4 discusses the data and related econometric issues and Section 5 presents the empirical results. The final section concludes the paper.

\section{Theoretical regularity}\label{section_theory}

Consumer demand economic theory assumes that consumer behavior is rational that consumers is in the pursuit of the maximum utility. Specifically, consumers will select a certain number of combinations of goods and services,  so that can maximize the utility he gets from the consumption of goods and services. Most demand system is derived from the utility maximization or cost minimization. According to the utility, cost function and the duality theory, such a demand system has some special properties ( adding-up, homogeneity, symmetry and negativity ) and mathematical expression is used to reflect the constraints and assumptions of demand system equation. The adding-up stems from budget constraint and requires the value of Marshallian demand or Hicks demand equals the total expenditure. Adding-up will be satisfied if $$\sum_{i=1}^N p_i q_i(y,p)=\sum_{i=1}^N p_i h_i(u,p)=y $$ or $$\sum_{i=1}^N w_i=1$$  where $ q_i(y,p)$ is the Marshallian demand function. $ h_i(u,p)$ is the Hicks demand function. $y$ is the total expenditure. $p$ is a vector of price and $w_i$ is the share of expenditure. Homogeneity is also stems from budget constraint and means that consumer decisions depends on the changes in real income and relative prices. Homogeneity will be satisfied if $$h_i(u,\lambda p)=h_i(u,p) \quad  q_i(\lambda y,\lambda p)=q_i(y,p)$$ Symmetry stems from cost function and requires cross-price derivative of Hicks demand function equal.$$\frac{\partial h_i(u,p)}{\partial p_j}=\frac{\partial h_j(u,p)}{\partial p_i}$$ Negativity stems from concave cost function and requires that the Hessian matrix of the cost function, $H$, is negative semidefinite. It also means that own-price elasticity of Hicks demand is not positive. $$\frac{\partial h_i(u,p)}{\partial p_i} \leq 0$$

In addition, as required by neoclassical microeconomics theory, the usefulness of flexible functional forms depends on whether they satisfy the theoretical regularity conditions of positivity, monotonicity, and curvature. In fact, as Barnett (2002) put it, without satisfaction all theoretical regularity conditions, 'the second-order conditions for optimizing behavior fail, and duality theory fails. The resulting first-order conditions, demand functions, and supply functions become invalid.' Positivity requires that the estimated cost be positive for all the data observations. More formally, positivity will be satisfied if the estimated cost is positive, $$\hat{C}(u,p)>0$$  Monotonicity requires that the first-order derivatives of the cost function be nonnegative. Monotonicity is satisfied if the estimated shares are positive, $$\hat{w_i}>0$$ Curvature requires that the cost function be a concave function of prices or, equivalently, that the Hessian matrix of the cost function, $H$ , be negative semidefinite. Curvature is satisfied if the Hessian matrix $$ H=\frac{\partial^{2}C(u,p)}{\partial p_i \partial p_j}$$  In summary, when we use the flexible functional forms to study consumer demand, we should test or impose the theoretical regularity conditions: positivity, monotonicity, curvature, adding-up, homogeneity and symmetry ( the negativity is equivalent to curvature ). It should be noted that regularity should not be treated as being equivalent to curvature alone, instead, it includes all the conditions above.

The theoretical regularity can be imposed or tested. In generally, positivity and monotonicity are satisfied automatically. Adding-up, homogeneity and symmetry are imposed before the parameters estimation. Curvature( or negativity) can be imposed or tested. In particular, Ryan and Wales (1998), drawing on related work by Lau (1978) and Diewert and Wales (1987), suggest a procedure for imposing local curvature conditions and apply their procedure to three locally flexible functional forms: the Almost Ideal Demand System (AIDS), the normalized quadratic (NQ), and the linear translog. Also, Moschini (1999) suggests a possible reparametrization of the basic translog of Christensen et al.(1975) to overcome some problems noted by Ryan and Wales(1998) and imposes curvature conditions locally in the basic translog. More recently, Serletis and Shahmoradi (2007) impose curvature conditions locally on the generalized Leontief model, and Chang and Serletis (2012) on the quadratic AIDS (QUAIDS). Also, Nana et al.(2014) impose curvature conditions on flexible functional forms for GNP functions approximated by a translog (TL) function. However, imposing the curvature condition more or less affect the flexibility of the model. A more prudent approach is to test it. In this paper, we test the positivity, monotonicity, curvature and impose adding-up, homogeneity and symmetry.

\section{Demand systems}\label{section_theory}

The demand systems approach allows us to estimate the consumer demand in a systems framework, which involves estimating the parameters of an aggregator function: a profit, expenditure or utility function. Let us assume that the consumer preferences is weak separability, which means that the estimated goods are a group separable from consumption goods. And then, the representative consumer has the following utility function:

\begin{equation} \label{eq:r1}
u= u(f(X),Z)
\end{equation}
where\emph{ X} is a vector of estimated goods, \emph{Z }is a vector of consumption goods outside the estimated goods and \emph{f(X)} is a aggregator function over estimated goods. At the same time, the consumer faces to a budget constrain:

\begin{equation} \label{eq:r2}
PX+QZ= m
\end{equation}
where \emph{P} refers to the vector of prices of the estimated goods, \emph{Q} refers to the vector of prices of the consumption goods outside the estimated goods, \emph{ m} is total expenditure. The assumption of weak separability means that

\begin{equation} \label{eq:r3}
\frac{\partial MRS_{ij}}{\partial z}=\partial (\frac{\partial u/\partial x_i}{\partial u/\partial x_j})/{\partial z}=0, \quad x_i,x_j\in X \quad and  \quad i\ne j
\end{equation}
That is, the marginal rate of substitution between any two components of \emph{ X} does not depend upon the values of  \emph{Z} , meaning that the demand for estimated goods is independent of relative prices outside the estimated goods.

Under the weak separability assumption, we will focus on the details of the demand for estimated goods, ignoring other types of goods, in the context of the following problem:

\begin{equation} \label{eq:r4}
\max f(X) \quad subject \ to \quad PX=y
\end{equation}
where \emph{X} is the vector of estimated goods. \emph{P} is the corresponding vector of prices, and y is the total expenditure on estimated goods.

The consumer demand system models can be divided into two types. The first type models derive from the specific algebraic function. The second type models are based on the utility function or cost function. In this section we briefly discuss three flexible functional forms. They are the Rotterdam model of  Theil (1965) and Barten (1964), belonging to the first type model, the Almost Ideal Demand System (AIDS) of Deaton and Muellbauer ( 1980a, 1980b ), belonging to the second type model, and the quadratic AIDS (QUAIDS) of Banks et al. (1997), belonging to the second type model.

\subsection{The Rotterdam model}

The Rotterdam model is first proposed by Theil(1965) and Barten(1964) and is called, after their domicile. Taylor Theorem is the basis of the Rotterdam model. They assume that the structure of demand models can be reflected by the First Order Approximation of the Taylor Theorem. Double-log function form is the jumping-off point of the model.

\begin{equation} \label{eq:r5}
\ln q_i=a_i+e_i\ln y+\sum_{k=1}^n e_{ik}\ln p_k
\end{equation}
where $q_i$ refers to the quantity of the goods. $p_k$ refers to the prices of the goods. \emph{y} refers to the total expenditure on goods. $e_i$ and  $e_{ik}$ are expenditure elasticity and price elasticity of Marshall demand, respectively. And then, totally differentiates the equation 5:

\begin{equation} \label{eq:r6}
\mathrm{d}\ln q_i=e_i\mathrm{d}\ln y+\sum_{k=1}^n e_{ik}\mathrm{d}\ln p_k
\end{equation}
According to the Slutsky function, equation 6 becomes:

\begin{equation} \label{eq:r7}
\mathrm{d}\ln q_i=e_i(\mathrm{d}\ln y-\sum_{k=1}^n w_k\mathrm{d}\ln p_k)+\sum_{k=1}^n \theta_{ik}\mathrm{d}\ln p_k
\end{equation}
where $\theta_{ik}$ indicates the price elasticity of Hicks demand. Then multiply each term by $w_i$ , which means the expenditure share of the $i^{th}$ good, and simplify:

\begin{equation} \label{eq:r8}
w_i\mathrm{d}\ln q_i=\alpha_i\mathrm{d}\ln Q+\sum_{k=1}^n \gamma_{ik}\mathrm{d}\ln p_k)
\end{equation}
where

\begin{equation} \label{eq:r9}
\mathrm{d}\ln Q=\mathrm{d}\ln y-\sum_{k=1}^n w_k\mathrm{d}\ln p_k=\sum_{k=1}^n w_k\mathrm{d}\ln q_k
\end{equation}

\begin{equation} \label{eq:r10}
\alpha_i=w_ie_i
\end{equation}

\begin{equation} \label{eq:r11}
\gamma_{ik}=w_i\theta_{ik}
\end{equation}

Some restriction from neoclassical demand theory can be imposed on the parameters of the model. The adding-up restriction is given by $\sum_{i=1}^n \alpha_i=1$ and $\sum_{i=1}^n \gamma_{ik}=0$. The homogeneity condition is imposed as $\sum_{i=1}^n \gamma_{ik}=0$. The Slutsky symmetry condition is given by $\gamma_{ik}=\gamma_{ki}$. However, the positivity, monotonicity and negativity need to be tested.  According to Deaton and Mullebauer(1980b), since all prices are positive, $\gamma$ will be negative semidefinite if and only if $\emph{S}$ is negative semidefinite. Consequently, symmetry and negativity of $\emph{S}$  can be tested by testing the same restrictions on $\gamma$. (Note,$s_{ij}=\frac{\partial h_i}{\partial p_j}=\frac{\partial q_i}{\partial p_j}+\frac{\partial q_i}{\partial x}q_j$) If the eigenvalues of the matrix $\gamma$ are all 0 or smaller, the model satisfy the negativity condition.

The expenditure elasticity is defined as:

\begin{equation} \label{eq:r11}
e_i=\alpha_i/w_i
\end{equation}
and the compensated price elasticity is given by:

\begin{equation} \label{eq:r11}
\theta_{ik}=\gamma_{ik}/w_i
\end{equation}

\subsection{The Almost Ideal Demand System}

The AIDS can be derived from the following indirect utility function:

\begin{equation} \label{eq:r11}
V(y,\mathbf{p})=(\ln y-\ln a(\mathbf{p}))/\beta_0 \prod_{k=1}^n p_k^{\beta_k}
\end{equation}
where\emph{ y} is the total expenditure,\emph{ k}=1,2,...,n means the number of goods,$\mathbf{p}$ is the vector of prices:

\begin{equation} \label{eq:r11}
\ln a(\mathbf{p})=\alpha_0+\sum_{k=1}^n \alpha_k \ln p_k+0.5\sum_{k=1}^n\sum_{j=1}^n \gamma_{kj}^*\ln p_k\ln p_j
\end{equation}
where $\alpha $ , $\beta $ and $\gamma $ are parameters. Solving for $\ln y $ yields the AIDS cost function, as in Deaton and Muellbauer (1980a,1980b):

\begin{equation} \label{eq:r11}
\ln c(u,\mathbf{p})=(1-u)\ln a(\mathbf{p})+u\ln b(\mathbf{p})
\end{equation}
and

\begin{equation} \label{eq:r11}
\ln b(\mathbf{p})=\ln a(\mathbf{p})+\beta_0\prod_{k=1}^n p_k^{\beta_k}
\end{equation}

Applying Roy's identity to the indirect utility function (14), or Shephard's lemma to the cost function
(16), we obtain the AIDS in budget share form:

\begin{equation} \label{eq:r11}
w_i=\alpha_i+\sum_{j=1}^n \gamma_{ij}\ln p_j+\beta_i\ln (\frac{y}{a(\mathbf{p})})
\end{equation}
where$ \gamma_{ij}=\frac{1}{2}(\gamma_{ij}^*+\gamma_{ji}^*)$

Economic theory imposes several restrictions on the parameters of the AIDS. In particular, symmetry requires $\gamma_{ij}=\gamma_{ji}$ for all \emph{i}, \emph{j}. The adding-up condition requires $\sum_{i=1}^n \alpha_i=1$, $\sum_{i=1}^n \gamma_{ij}=0$ and $\sum_{i=1}^n \beta_i=0$. The homogeneity condition is imposed as $\sum_{j=1}^n \gamma_{ij}=0$. However, the positivity, monotonicity and negativity also need to be tested. Deaton and Muellbauer (1980a, 1980b) suggest that use $k_{ij}=p_ip_js_{ij}/x$ to test the negativity. The specific value is: $k_{ij}=\gamma_{ij}+\beta_i\beta_j\ln(\frac{y}{a(p)})-w_i\delta_{ij}+w_iw_j$, where $\delta=1$ if $i=j$, and 0 otherwise. If the eigenvalues of the matrix $\emph{K}$ are all 0 or smaller, the model satisfy the negativity condition.

The expenditure elasticity is defined as:

\begin{equation} \label{eq:r11}
e_i=1+\frac{\beta_i}{w_i}
\end{equation}
and the uncompensated price elasticity is given by:

\begin{equation} \label{eq:r11}
e_{ij}=-\delta_{ij}+w_i^{-1}(\gamma_{ij}-\beta_i(\alpha_i+\sum_{j=1}^n \gamma_{ij}\ln p_j))
\end{equation}

\subsection{The Quadratic Almost Ideal Demand System}

According to Banks \emph{et al.} (1997), the QUAIDS has the following indirect utility function:

\begin{equation} \label{eq:r11}
\ln v(\mathbf{p},y)=\left\{\left[(\frac{\ln y-\ln a(\mathbf{p})}{b(\mathbf{p})})\right]^{-1}+ \lambda(\mathbf{p})\right\}^{-1}
\end{equation}
where\emph{ y} is total expenditure, $\mathbf{p}$ is the vector of prices, $a(\mathbf{p})$ is a differentiable, homogeneous function of degree one in prices, and $\lambda(\mathbf{p})$ and $b(\mathbf{p})$ are differentiable, homogeneous functions of degree zero in prices. In fact, $\lambda(\mathbf{p})$ takes the form $$\lambda(\mathbf{p})=\sum_{k=1}^n \lambda_k\ln p_k \quad where \ \sum_{k=1}^n \lambda_k=0$$
and $\emph{k}=1,2,...,n$ denotes the number of goods.

The specifications of $ a(\mathbf{p})$ and $ b(\mathbf{p})$ in (21) are similar to those in the AIDS of Deaton and Muellbauer (1980a,1980b). They are sufficiently flexible to represent any arbitrary set of first and second derivatives of the cost function, as follows: $$\ln a(\mathbf{p})=\alpha_0+\sum_{k=1}^n \alpha_k \ln p_k+0.5\sum_{k=1}^n\sum_{j=1}^n \gamma_{kj}^*\ln p_k\ln p_j$$ $$\ln b(\mathbf{p})=\beta_0\prod_{k=1}^n p_k^{\beta_k}$$

Solving for $\ln y$, yields the QUAIDS cost function

\begin{equation} \label{eq:r11}
\ln C\equiv\ln y=\ln a(\mathbf{p})+\frac{b(\mathbf{p})\ln v(\mathbf{p},y)}{1-\lambda(\mathbf{p})\ln v(\mathbf{p},y)}
\end{equation}

Applying Shephard's lemma to the cost function (22) or Roy's identity to the indirect utility function (21) yields the QUAIDS share equation:

\begin{equation} \label{eq:r11}
w_i=\alpha_i+\sum_{j=1}^n \gamma_{ij}\ln p_j+\beta_i\ln (\frac{y}{a(\mathbf{p})})+\frac{\lambda_i}{b(\mathbf{p})}\left[\ln (\frac{y}{a(\mathbf{p})}\right]^2
\end{equation}
where $w_i$ is the \emph{ i}th budget share and  $\alpha $ , $\beta $ , $\gamma $ and $\lambda $ are parameters. By setting $\lambda_i=0(i=1,2,...,n) $,equation (23) reduces to the AIDS share equation.

Economic theory imposes several restrictions on the parameters of the model. In particular, symmetry requires $\gamma_{ij}=\gamma_{ji}$ for all \emph{i}, \emph{j}. The adding-up condition requires $\sum_{i=1}^n \alpha_i=1$, $\sum_{i=1}^n \gamma_{ij}=0$ , $\sum_{i=1}^n \beta_i=0$ and $\sum_{i=1}^n \lambda_i=0$. The homogeneity condition can be imposed as $\sum_{j=1}^n \gamma_{ij}=0$. However, the positivity, monotonicity and negativity also need to be tested. Banks, Blundell and Lewbel (1997) suggest that assess the negativity conditions by examining the matrix with elements $w_i\left[e_{ij}^c\right]$, which should be symmetric and negative semidefinite in general, where $e_{ij}^c$ is the compensated price elasticity. If the eigenvalues of the matrix are all 0 or smaller, the model satisfy the negativity condition.

The expenditure elasticity is defined as:

\begin{equation} \label{eq:r11}
e_i=1+w_i^{-1}\left[\beta_i+\frac{2\lambda_i}{b(\mathbf{p})}\ln (\frac{y}{a(\mathbf{p})})\right]
\end{equation}
and the uncompensated price elasticity is given by:

\begin{equation} \label{eq:r11}
e_{ij}=w_i^{-1}\left\{\gamma_{ij}-(\beta_i+\frac{2\lambda_i}{b(\mathbf{p})})\left[\ln (\frac{y}{a(\mathbf{p})})\right](\alpha_j+\sum_{k=1}^n\gamma_{jk}\ln (p_k))-\frac{\lambda_i\beta_j}{b(\mathbf{p})}\left[\ln (\frac{y}{a(\mathbf{p})})\right]^2\right\}-\delta_{ij}
\end{equation}
where $\delta=1$ if $i=j$, and 0 otherwise.

\section{Econometric issues and data}\label{section_theory}

In this section,we use three flexible functional forms, the Rotterdam, AIDS and QUAIDS model to investigate the demand for road transport in China in a systems framework. We pay more attention to theoretical regularity and argue that, unless regularity is attained by luck, flexible functional forms should always be estimated subject to regularity and all regularity conditions are equally important. In doing so, we impose the adding-up, homogeneity and symmetry conditions on the three models and test the positivity, monotonicity and negativity conditions. And then, select the appropriate model. Only in this way, the result of parameter and elasticities estimation is meaningful and can produce inference consistent with neoclassical microeconomic theory.

Firstly, considering the availability of data, we define the transportation goods in this paper. The private transportation includes Mini passenger vehicles and small passenger vehicles. The local transportation includes Public buses and taxi. The intercity transportation includes medium passenger vehicles and large passenger vehicles. And our underlying assumption is that individuals with similar geographical characteristics have similar decision-making behaviors on their expenditures. Thus we use the panel data in 31 provinces in mainland China from 2002 to 2014. We use the following function to calculate annual expenditure data.

\begin{equation} \label{eq:r11}
y=Q\times VMT\times FE\times P
\end{equation}
where \emph{ Q} refers to the quantity of the vehicle and the panel data are from the National Bureau of Statistics of China. \emph{VMT} means the annual average vehicle mileage traveled, which is closely related to the economic growth level and road traffic infrastructure (Huo et al.,2012a), and some studies and estimates on China's VMT data can be seen in He et al.(2005) and Huo et al.(2012a). Considering the regional differences and availability of the data, we use the national average VMT data in 2002 as baseline data. Then, by using the 31 provincial average passenger transport distances, we estimate the provincial annual average VMT data for each transportation goods from 2002 to 2014. Our VMT estimates are supported by many survey papers and empirical studies in existing literatures (please see table 2), indicating that our estimated results are reasonably acceptable. \emph{ FE} denotes the fuel economy and the data are from the previous literatures (please see table 2). \emph{P} means the China's provincial gasoline and diesel annual average retail prices and the data are from the journal China Petroleum and Chemical Industry and the journal International Petroleum Economics.

The associated prices for these expenditure categories are normalized (with last year=100) annual consumer price indices from the National Bureau of Statistics of China.

\begin{table}[h!]\small
\setlength{\abovecaptionskip}{0pt}
\setlength{\belowcaptionskip}{10pt}
\centering
\begin{threeparttable}[b]
\captionsetup{font={small},labelfont=bf,textfont=normalfont,justification=raggedright,labelsep=space,labelsep=newline,singlelinecheck=off}
\caption{The annual average VMT and fuel economy} \label{tab1}
\footnotesize
\begin{tabular}{lcccccc}
\hline
             & \multicolumn{2}{c}{Private transportation ($x_1$)} & \multicolumn{2}{c}{Local transportation($x_2$)} & \multicolumn{2}{c}{intercity transportation($x_3$)} \\ \cline{2-4} \cline{5-7}
Parameters   & MNPV & SPV  & PB & Taxi & MPV & LPV\\
\hline

Baseline VMT\tnote{a}       & 34 & 33.6 & 57.2 & 74.9 & 47.3 & 48.6  \\
Fuel economy\tnote{b}       & 6.38 & 9 & 33 & 8.7 & 25.97 & 32.6 \\
\hline

\end{tabular}
VMT: The annual average vehicle mileage traveled;MNPV:Mini passenger vehicles;SPV:Small passenger vehicles;PB:Public buses;MPV:Medium passenger vehicles;LPV:Large passenger vehicles
\begin{tablenotes}
   \item[a] Unit:1000 kilometers. Baseline VMT refer to the 2002 national average VMT. The VMT of MNPV, SPV, MPV and LPV see China Energy Databook v.7.0, October 2008. The VMT of PB and Taxi see Huo \emph{et al.}(2012a).
   \item[b] Unit:L/100km. The fuel economy:MNPV (Wang \emph{et al.}(2005)),SPV and MPV (Huo\emph{ et al.}(2012b)),LPV (Zhang \emph{et al.}(2014)),PB (Shen \emph{et al}.(2014)),Taxi (Hu \emph{et al.}(2012)).

\end{tablenotes}
\end{threeparttable}
\end{table}

\section{Empirical evidence}\label{section_theory}

In this study, we use three flexible function forms, the Rotterdam model, the AIDS and QUAIDS model to investigate the demand for road transportation in a systems framework. We pay explicit attention to theoretical regularity and argue that, unless regularity is attained by luck, flexible functional forms should always be estimated subject to regularity and all regularity conditions are equally important.

\subsection{Parameter estimates}

We estimate the parameters by using Zellner's seemingly unrelated regression in the Rotterdam model and nonlinear seemingly unrelated regression in the AIDS and QUAIDS model. In addition, we impose the adding-up, homogeneity and symmetry condition on the parameters of the three models, and then, test the negativity,   positivity and monotonicity conditions, considering that imposing the conditions in advance would restrict the flexibility of the models. Finally, since the budget shares sum to 1, the disturbance covariance matrix is singular. To address the issue, we follow Barten(1969) and drop the last equation in each model. Table 3 contain a summary of results from the Rotterdam, the AIDS and QUAIDS models in terms of parameter estimates.

\begin{table}[h!]\small
\setlength{\abovecaptionskip}{0pt}
\setlength{\belowcaptionskip}{10pt}
\centering
\begin{threeparttable}[b]
\captionsetup{font={small},labelfont=bf,textfont=normalfont,justification=raggedright,labelsep=space,labelsep=newline,singlelinecheck=off}
\caption{The parameter estimates} \label{tab1}
\small
\begin{tabular}{lccc}
\hline
Parameters   & The Rotterdam model & AIDS  & QUAIDS \\
\hline
$\alpha_1$     &0.880(0.006) &1.879(0.048)  &0.406(0.014) \\
$\alpha_2$     &0.030(0.002) &-0.278(0.018)  &0.182(0.006) \\
$\alpha_3$    &0.090(0.005) &-0.601(0.035)  &0.412(0.011) \\
$\gamma_{11}$     &-0.430(0.099) &0.330(0.231)  &-0.143(0.222) \\
$\gamma_{21}$     &0.118(0.026) &-0.125(0.089)  &0.029(0.087) \\
$\gamma_{31}$     &0.312(0.083) &-0.206(0.170)  &0.115(0.165) \\
$\gamma_{22}$     &-0.213(0.031) &0.090(0.093)  &0.037(0.093) \\
$\gamma_{32}$     &0.095(0.034) &0.035(0.098)  &-0.066(0.098) \\
$\gamma_{33}$     &-0.407(0.077) &0.171(0.159)  &-0.049(0.156) \\
$\beta_1$         &              &0.211(0.008)  &0.131(0.009) \\
$\beta_2$         &              &-0.066(0.003)  &-0.041(0.003) \\
$\beta_3$         &              &-0.145(0.006)  &-0.090(0.007) \\
$\lambda_1$         &              &             &-0.007(0.001) \\
$\lambda_2$         &              &             &0.002(0.001) \\
$\lambda_3$         &              &             &0.005(0.001) \\
Positivity violation&  0            &  0           &0 \\
Monotonicity violation&  0            &  0           &0 \\
\hline

\end{tabular}

$\emph{Note:}$ Goods 1=Private transportation 2= local transportation 3= intercity transportation.Sample period,2002-2014.Numbers in parentheses are standard error.
\end{threeparttable}
\end{table}

\begin{landscape}
\begin{table}[h!]\small
\setlength{\abovecaptionskip}{0pt}
\setlength{\belowcaptionskip}{10pt}
\centering
\begin{threeparttable}[b]
\captionsetup{font={small},labelfont=bf,textfont=normalfont,justification=raggedright,labelsep=space,labelsep=newline,singlelinecheck=off}
\caption{Expenditure and price elasticities} \label{tab1}
\scriptsize
\begin{tabular}{lccccccccc}
\hline
Goods   & Model & \multicolumn{8}{c}{Elasticities}\\
\cline{3-10}
&   & Expenditure elasticity& \multicolumn{7}{c}{Price elasticity}\\
\cline{4-10}
 &   &   & \multicolumn{3}{c}{Uncompensated own- and cross-price} &  & \multicolumn{3}{c}{Compensated own- and cross-price} \\
\cline{4-6} \cline{8-10}

        &       &$e_i $  &$ e_{i1}^u $  &$ e_{i2}^u$  &$e_{i3}^u$ & &$e_{i1}^c$  &$e_{i2}^c$  &$e_{i3}^c$\\
 \hline
Private transportation &  &  &  &  &  &  &  &  & \\
          & Rotterdam   &1.261(0.009)  &-1.496(0.141)  &0.056(0.037)  &0.179(0.119) & &-0.616(0.141)  &0.169(0.037) &0.447(0.119) \\
          & AIDS   &1.303(0.012)  &-1.097(0.325)  &-0.094(0.126)  &-0.112(0.241)&  &-0.187(0.325)  &0.023(0.126)  &0.164(0.241) \\
          & QUAIDS   &1.133(0.005)  &-1.249(0.318)  &0.014(0.125)  &0.102(0.236)&  &-0.458(0.318)  &0.115(0.125)  &0.343(0.236) \\
Local  transportation &  &  &  &  &  &  &  &  & \\
         & Rotterdam   &0.338(0.018)  &1.309(0.285)  &-2.401(0.343)  &0.985(0.373) & &1.313(0.285)  &-2.370(0.343)  &1.057(0.373) \\
         & AIDS   &0.267(0.037)  &-0.005(0.979)  &-0.206(1.035)  &-0.056(1.091) & &0.181(0.981)  &-0.182(1.035)  &0.001(1.091) \\
          & QUAIDS   &0.678(0.016)  &0.423(0.966)  &-0.520(1.037)  &-0.581(1.085)&  &0.896(0.966)  &-0.459(1.037)  &-0.437(1.085) \\
Intercity  transportation &  &  &  &  &  &  &  & &  \\
         & Rotterdam   &0.423(0.025)  &1.175(0.387)  &0.410(0.158)  &-2.007(0.362)&  &1.470(0.390)  &0.448(0.158)  &-1.918(0.362) \\
         & AIDS   &0.315(0.030)  &0.321(0.791)  &-0.028(0.463)  &-0.607(0.742) & &0.540(0.791)  &0.001(0.463)  &-0.540(0.742) \\
          & QUAIDS   &0.699(0.013)  &0.640(0.776)  &-0.248(0.460)  &-1.091(0.734) & &1.128(0.776)  &-0.185(0.460)  &-0.943(0.734) \\

\hline

\end{tabular}

$\emph{Note:}$ Goods 1=Private transportation 2= local transportation 3= intercity transportation.Sample period,2002-2014.Numbers in parentheses are standard error.
\end{threeparttable}
\end{table}

\end{landscape}
Because regularity has not been attained (by luck), we estimate the models by imposing adding-up, homogeneity and symmetry condition using the methodology discussed in Section 3. As can be seen in Table 3, the results satisfy the three conditions, but the negativity, positivity and monotonicity would be tested. And then, the standard error indicates that the parameters estimating are significant.

\subsection{Road transportation demand elasticities}

In the demand systems approach to estimation of economic relationships, the primary interest, especially in policy analysis, is in how the arguments of the underlying function affect the quantities demanded. This is conventionally completely expressed in terms of expenditure and price elasticities.

We report the expenditure and own- and cross-price elasticities (evaluated at the mean of the data) in Table 4 with standard error in parentheses, using the Delta method to calculate. All road transportation demand elasticities are estimated on the basis of parameter estimates and sample means of explanatory variables.

In Table 4, we show the expenditure elasticities, the uncompensated own- and cross-price elasticities and compensated own- and cross-price elasticities for each of the three transportation goods,using three models. At this time, we do not know whether these three models meet the negativity condition. So, we take the negativity test for these three models, using the methods in Section 3. For the Rotterdam model, the matrix $\gamma$ is given by:

\begin{equation}       
\left[                 
  \begin{array}{ccc}   
    -0.4298106      &  & \\  
    0.1180672 & -0.2130839 & \\
   0.3117434 & 0.0950168 & -0.4067602 \\
  \end{array}
\right]                 
\end{equation}
And the eigenvalues of the matrix $\gamma$ are 3.333e-08, -0.31865725 and -0.73099749. The first eigenvalue is greater than 0. Therefore, the Rotterdam model does not satisfy the negativity condition, indicating that the results of parameters and elasticities in the Rotterdam model become invalid.

For the AIDS model, considering $k_{ij}=\gamma_{ij}+\beta_i\beta_j\ln(\frac{y}{a(p)})-w_i\delta_{ij}+w_iw_j$, the matrix $\emph{K}$ is given by:

\begin{equation}       
\left[                 
  \begin{array}{ccc}   
    -0.22078782      &  & \\  
    0.04609013  & -0.09179885 & \\
    0.17469769 &  0.04570872 & -0.22040642 \\
  \end{array}
\right]                 
\end{equation}
And the eigenvalues of the matrix $\emph{K}$ are -1.110e-16, -0.13769785 and -0.39529524. The three eigenvalues are all less than 0, implying that the matrix is negative and the AIDS satisfy the negativity condition. This also means that the results of parameters and elasticities in the AIDS model start to make sense.

For the QUAIDS model, the matrix with elements $w_i\left[e_{ij}^c\right]$ is given by:

\begin{equation}       
\left[                 
  \begin{array}{ccc}   
    -0.31979755      &  & \\  
    0.08052734  & -0.04122938 & \\
    0.23927301 &  -0.03928298 & -0.19998905 \\
  \end{array}
\right]                 
\end{equation}
And the eigenvalues of the matrix are 6.254e-06, -0.0384866 and -0.52253564. The first eigenvalue is greater than 0. Therefore, the QUAIDS model does not satisfy the negativity condition, indicating that the results of parameters and elasticities in the QUAIDS model become invalid.

In this paper, the adding-up, homogeneity and symmetry conditions are imposed in all models, implying that all models satisfy the three conditions. And then, we test the positivity, monotonicity and negativity condition of the three models, finding that the three models are all satisfy the positivity and monotonicity, but only the AIDS model satisfy the negativity condition (see Table 5). This means that only the AIDS model satisfy all theoretical regularity conditions, indicating that the results of parameters and elasticities in the AIDS model make sense. \footnote{However, as can be seen, the standard error of the price elasticity is not significant. This may derive from the quality of the data. Due to data availability, we use the panel data in province. Compared with the micro-survey data, the aggregate data of province makes the heterogeneity among consumers fuzzy. But, the theoretical regularity (adding-up, homogeneity, symmetry, negativity,  positivity and monotonicity ) is the basis of the models. In this sense, the regularity is more important than the significance. Besides, the results of the elasticities is consistent with the economic practice. Above all, the AIDS model is the most suitable model to study the demand for road transport in China and the results of the AIDS model are meaningful. In the future, we will use the micro data to carry out the corresponding research, so that the results will become better.} And the results of AIDS elasticities estimates are reported in the Table 6.

\begin{table}[h!]\small
\setlength{\abovecaptionskip}{0pt}
\setlength{\belowcaptionskip}{10pt}
\centering
\begin{threeparttable}[b]
\captionsetup{font={small},labelfont=bf,textfont=normalfont,justification=raggedright,labelsep=space,labelsep=newline,singlelinecheck=off}
\caption{The theoretical regularity of three models} \label{tab1}
\scriptsize
\begin{tabular}{lcccccc}
\hline
Regularity   & Positivity &  Monotonicity  & Curvature&Adding-up  & Homogeneity&Symmetry  \\
\hline
The Rotterdam model     & $\surd$ & $\surd$   &$\times$  & $\surd$ & $\surd$ &$\surd$ \\
 AIDS   & $\surd$ & $\surd$   &$\surd$  & $\surd$ & $\surd$ &$\surd$ \\
QUAIDS   & $\surd$ & $\surd$   &$\times$  & $\surd$ & $\surd$ &$\surd$ \\

\hline

\end{tabular}
\end{threeparttable}
\end{table}

\begin{table}[h!]\small
\setlength{\abovecaptionskip}{0pt}
\setlength{\belowcaptionskip}{10pt}
\centering
\begin{threeparttable}[b]
\captionsetup{font={small},labelfont=bf,textfont=normalfont,justification=raggedright,labelsep=space,labelsep=newline,singlelinecheck=off}
\caption{AIDS expenditure and price elasticities} \label{tab1}
\scriptsize
\begin{tabular}{lcccccccc}
\hline

 Goods & Expenditure elasticity& \multicolumn{7}{c}{Price elasticity}\\
\cline{3-9}
 &   & \multicolumn{3}{c}{Uncompensated own- and cross-price} &  & \multicolumn{3}{c}{Compensated own- and cross-price} \\
\cline{3-5} \cline{7-9}

    &$e_i $  &$ e_{i1}^u $  &$ e_{i2}^u$  &$e_{i3}^u$ & &$e_{i1}^c$  &$e_{i2}^c$  &$e_{i3}^c$\\
 \hline
(1)  &  &  &  &  &  &  &  & \\

            &1.303  &-1.097  &-0.094  &-0.112 &  &-0.187  &0.023  &0.164 \\
            &(0.012)  &(0.325)  &(0.126)  &(0.241)&  &(0.325)  &(0.126)  &(0.241) \\
(2)&  &  &  &  &  &  &  & \\

          &0.267  &-0.005  &-0.206  &-0.056 & &0.181  &-0.182  &0.001 \\
          &(0.037)  &(0.979)  &(1.035)  &(1.091) & &(0.981)  &(1.035)  &(1.091) \\
(3)  &  &  &  &  &  &  & &  \\

            &0.315  &0.321  &-0.028  &-0.607 & &0.540  &0.001  &-0.540 \\
            &(0.030)  &(0.791)  &(0.463)  &(0.742) & &(0.791)  &(0.463)  &(0.742) \\

\hline

\end{tabular}

$\emph{Note:}$ Goods (1)=Private transportation (2)= local transportation (3)= intercity transportation.Sample period,2002-2014.Numbers in parentheses are standard error.
\end{threeparttable}
\end{table}

 As expected, all the expenditure elasticities, $e_1$, $e_2$, and $e_3$, are positive, implying that private transportation, local transportation and intercity transportation are all normal goods for consumers. Especially, the value of $e_1$ is greater than 1, implying that the private transportation is a luxury among the transportation goods. According to the Traffic Management Bureau in the Ministry of Public Security, by the end of 2015, the national average of 31 per 100 households have private cars. The rate of private cars is relatively low and most people can not afford the private cars. Correspondingly, the value of $e_2$ and $e_3$ is larger than 0 and smaller than 1, which means that the local and intercity transportation are all necessities. National Bureau of Statistics of China data show that since 2003, the investment in road transportation increased by 6 times to 2451.316 billion RMB yuan. More importantly, the government continue to increase investment in public transport infrastructure. As a result, the total road mileages and public transport operating lines of China witnessed rapid growths, and increased from 3.457 million and 0.126 million km (in 2006) to 4.464 million and 0.620 million km(in 2014), respectively. All of this provide convenient conditions for the local and intercity transportation of the residents to travel.

 The own-price elasticities($e_{ii}^u$ or $e_{ii}^c$) are all negative(as predicted by the theory), with their compensated own-price elasticities being less than 1, which indicates that the demand for all three types of transportation are inelastic for the consumers. However, the uncompensated own-price elasticities are all greater than the compensated own-price elasticities. This is because the uncompensated price elasticities describe the total effect of price change and the compensated price elasticities only describe the substitution effect. In particular, the value of $e_{11}^u$ is greater than 1, which means that the demand for the private transportation is elastic. As can be seen in the Table 5, the absolute value of the uncompensated own-price elasticity of the private transportation is the largest among all three types of transport goods, along with the uncompensated own-price of the local transportation has the smallest value. In general, the cost of the local transportation is lowest among all three types of transport goods, accounting for a relatively small proportion of daily expenses. As a result, the residents are relatively insensitive to changes in prices. On the contrary, the private transportation is relatively expensive and need to burden some additional costs, for example, the costs of traffic jams, which account for a relatively large proportion of daily expenses. Therefore, when rise the fuel prices or tax, there will be a significant reduction of demand of the private transportation.

 According to Nicholson(2011), the homogeneity and the negativity of the demand function imply that, any kind of items, at least, is a net substitute for another kind of items. The net substitute is defined that the rising in price of goods $\emph{k}$ leads to the decline of the compensated demand of goods $\emph{j}$. Namely, the goods $\emph{k}$ and goods $\emph{j}$ is a net substitute. And the the gross substitute and complement mainly depends on the uncompensated price elasticities. One undesirable characteristic of the gross definitions of substitutes and complements is that they are not symmetric. It is possible for $x_1$ to be a substitute for $x_2$ and at the same time for $x_2$ to be a complement of $x_1$.

 Regarding the uncompensated cross-price elasticities ($e_{ij}^u$ ), we see that all of that are negative except the $e_{31}^u$, implying that the private and the local transportation, the local and intercity transportation are gross complements. Note that the $e_{13}^u$ is also negative, which reflect the asymmetry of the gross definitions. This means that the private transportation to be a substitute for the intercity transportation and at the same time for the intercity transportation to be a complement of the private transportation. With the development of economy and the improvement of living standard and urbanization level, the private transportation has become a fashion choice for residents to travel. What's more, a considerable part of the consumers of the intercity traffic is the migrant workers, who may be more likely to choose the private transportation for the convenience when they return hometown. And there may be a part of psychological factors to prompt them to choose the private transportation. In addition, short-haul travelers or business people may be more inclined to choose the private transportation, considering the comfort, convenience and efficiency of the private transportation. However, a part of the people choose the intercity transportation due to a variety of reasons, for example, big cost of the private transportation, without driver's licences and so on.

 Finally, given the nature of our data, it should be noted that the obtained elasticities represent long-run responses rather than short-run behavior.

\section{Conclusion}\label{section_theory}
Most published studies that use flexible functional forms have ignored the theoretical regularity conditions implied by microeconomic theories. Moreover, even a few studies have checked and/or imposed regularity conditions, most of them equate curvature alone with regularity, thus ignoring or minimizing the importance of monotonicity, positivity, adding-up, homogeneity and symmetry. We argue that without satisfaction all theoretical regularity conditions (positivity, monotonicity, curvature,adding-up, homogeneity and symmetry), the resulting inferences are questionable, since violations of regularity conditions invalidate the duality theory on which economic functions are based. We believe that unless regularity is attained by luck, flexible functional forms should always be estimated subject to regularity and all regularity conditions are equally important.

In this paper, we provide an empirical selection of three most widely used flexible demand system functional forms: The Rotterdam model, AIDS and QUAIDS model. We pay explicit attention to theoretical regularity. And then, apply the suitable to investigate the demand for road transportation in China using the panel data in 31 provinces in mainland China from 2002 to 2014, over a 13-year period, on three expenditure categories in the transportation goods: the private, local and the intercity transportation. In doing so, We impose the adding-up, homogeneity and symmetry conditions on the three models,and then, test the negativity,  monotonicity and positivity conditions, in order to protect the flexibility of the models. Our findings in terms of negativity are disappointing in the case of the Rotterdam and QUAIDS models. The results of parameters and elasticities in the Rotterdam and QUAIDS models become invalid. The empirical results show that the AIDS is the only model that is able to provide inferences about the demand for road transportation in China that are consistent with neoclassical microeconomic theory.

Based on the AIDS model, our estimates of the expenditure elasticities are positive. And the value of $e_1$ is greater than 1, implying that the private transportation is a luxury among the transportation goods. The own-price elasticities are all negative. In particular, the value of $e_{11}^u$ is greater than 1, which means that the residents are relatively sensitive to changes in prices of the private transportation. The result of the cross-elasticities show that the private and the local transportation, the local and intercity transportation are gross complements. However, the private transportation is a substitute for the intercity transportation, while the intercity transportation is a complement of the private transportation.

Based on our evidence, in order to achieve the China's commitment of greenhouse gases emissions to peak by 2030, in the transport field, the China government should make different policies according to the transport goods. In concrete terms, the expenditure and price elasticities of demand for the private transportation are all greater than 1, implying that the magnitude of the change in demand is greater than the magnitude of changes in expenditure and price changes with other conditions unchanged. This means that reducing spending or raising price can effectively reduce demand for the private transportation, having a favorable impact on environment and residents' health (for example, individual income tax on personal income, Restrictions for purchasing or driving, fuel tax on the private vehicles, and so on). However, it will cause some negative impact on the economy. Therefore, the government should weigh its advantages against its disadvantages before the policy is made. It is a remarkable fact that price and expenditure of the demand for the local and intercity transportation are all inelastic, indicating that the consumer of the local and intercity transportation is not sensitive to price or expenditure changes. The results show that raising price in the local or intercity transportation does not effectively reduce demand. For example, the public transport in Beijing raised price in 2014. The measures improve the residents' travel expenses, but does not reduce the demand. This may affect the welfare of the residents to a certain extent. In summary, the traffic policy should be adjusted according to transport goods. Policy makers need to set clear goals and weigh the pros and cons before making a policy.


\end{onehalfspace}

\clearpage
\newpage
\section*{References}

{\footnotesize

Aepli, M.(2014).Consumer demand for alcoholic beverages in Switzerland: a two-stage quadratic almost ideal demand system for low, moderate, and heavy drinking households. Agricultural and Food Economics, 2, 1--27.

Angulo, A., Atwi, et al. (2014). Economic analysis of the water demand in the hotels and restaurants sector: Shadow prices and elasticities. Water Resources Research, 50, 6577--6591.

Banks J, Blundell R, Lewbel A. (1997). Quadratic Engel curves and consumer demand. The Review of Economics and Statistics, 79, 527--539.

Barten, A.P. (1964). Consumer demand functions under conditions of almost additive preferences.Econometrica, 32, 1--38.

Barten AP.(1969). Maximum likelihood estimation of a complete system of demand equations. European Economic Review, 1, 7--73.

Barnett WA.(2002). Tastes and technology: curvature is not sufficient for regularity. Journal of Econometrics, 108, 199--202.

Byron, R.P.(1970). A Simple Method For Estimating Demand Systems Under Separable Utility Assumptions. Review of Economic Studies, 37, 261--274.

Chang D, Serletis A.(2012). Imposing local curvature in the QUAIDS. Economics Letters, 115, 41-- 43.

Chang, D., Serletis, A.(2014). The demand for gasoline: evidence from household survey data. Journal of Applied Econometrics, 29, 291--313.

Chen, S.M., He, L.Y.(2014). Welfare loss of China's air pollution: How to make personal vehicle transportation policy. China Economic Review, 31, 106--118.

Christensen LR, Jorgenson DW, Lau LJ.(1975). Transcendental logarithmic utility functions. American Economic Review, 65, 367--383.

Dargay, J. and Vythoulkas, P.(1999). Estimation of a dynamic car ownership model: a pseudo-panel approach, Journal of Transport Economics and Policy, 33, 287--302.

Deaton , A.(1978). Specification and Testing in Applied Demand Analysis. The Economic Journal, 88, 524--536.

Deaton A, Muellbauer J. (1980a). An almost ideal demand system. American Economic Review, 70, 312--326.

Deaton A, Muellbauer J. (1980b). Economics and Consumer Behavior. Cambridge University Press, Cambridge, Chapter 3.

Diewert WE, Wales TJ. (1987). Flexible functional forms and global curvature conditions. Econometrica, 55, 43--68.

Fashogbon, A. E., Oni, O. A.(2013). Heterogeneity in Rural Household Food Demand and Its Determinants in Ondo State, Nigeria: An Application of Quadratic Almost Ideal Demand System. Journal of Agricultural Science, 5, 169-177.

Ghalwash TM.(2008). Demand for environmental quality: an empirical analysis of consumer behavior in Sweden. Environmental and Resource Economics, 41, 71--87.

Graham, D.J., Glaister, S.(2002). The demand for automobile fuel: a survey of elasticities, Journal of Transport Economics and Policy, 36, 1--26.

Graham, D.J., Glaister, S.(2004). Road traffic demand elasticities: a review. Transport Reviews, 24, 261--274.

Guo, X.R., Cheng, S.Y., Chen, D.S., Zhou, Y., Wang, H.Y. (2010). Estimation of economic costs of particulate air pollution from road transport in China. Atmospheric Environment, 44, 3369--3377.

Hanly, M., Dargay, J. and Goodwin, P. (2002). Review of Income and Price Elasticities in the Demand for Road Traffic (London: Department for Transport).

Hao, J., Wang, L.(2005). Improving urban air quality in China: Beijing case study. J. Air Waste Manage. Assoc, 55, 1298--1305.

He, L.Y., Chen, Y.(2013). Thou shalt drive electric and hybrid vehicles: Scenario analysis on energy saving and emission mitigation for road transportation sector in China. Transport Policy, 25, 30--40.

He, KB., Huo, H., Zhang, Q., He, DQ., An, F.,Wang, M., Walsh, MP.(2005). Oil consumption and CO2 emissions in china's road transport: current status, future trends, and policy implications.Energy Policy, 33, 1499--1507.

Hu, J., Wu, Y., Wang, Z., Li, Y., Zhou, Y., Wang, H., Bao, X., Hao, J.(2012). Real-world fuel efficiency and exhaust emissions of light-duty diesel vehicles and their correlation with road conditions. Journal of Environmental Sciences, 24, 865--874.

Huo, H., Zhang, Q., He, K., Yao, Z., Wang, M.(2012a). Vehicle-use intensity in China: current status and future trend. Energy Policy, 43, 6--16.

Huo, H., He, K., Wang, M., Yao, Z.(2012b). Vehicle technologies, fuel-economy policies, and fuel-consumption rates of Chinese vehicles. Energy Policy, 43, 30--36.

Iootty, M., Helder Pinto Jr., Ebeling, F.(2009). Automotive fuel consumption in Brazil: Applying static and dynamic systems of demand equations. Energy Policy, 37, 5326-5333.

Lau LJ. (1978). Testing and imposing monotonicity, convexity, and quasi-convexity constraints. In Production Economics: A Dual Approach to Theory and Applications, Fuss M, McFadden D (eds). North-Holland: Amsterdam, 409--453.

Lewis, K. and Widup, D.(1982). Deregulation and rail truck competition: evidence from a translog transport demand model, Journal of Transport Economics and Policy, 16, 139--149.

Moschini G. (1999). Imposing local curvature in flexible demand systems. Journal of Business and Economic Statistics, 17, 487--490.

Nana, G. C., Larue, B.(2014). Imposing curvature conditions on flexible functional forms for GNP functions. Empirical Economics, 47, 1411--1440.

Nicholson, W.(2011). Microeconomic Theory Basic Principles and Extensions. South-Western College Pub.

Oum, T. (1979). Derived demand for freight transport and inter-modal competition in Canada, Journal of Transport Economics and Policy, 13, 149--168.

Petr, J. (2014). Consumer Demand System Estimation and Value Added Tax Reforms in the Czech Republic.Finance a Uver: Czech Journal of Economics and Finance, 64, 246--273.

Romilly, P., Song, H. and Liu, X.(1998). Modelling and forecasting car ownership in Britain, Journal of Transport Economics and Policy, 32, 165--185.

Ryan DL, Wales TJ.(1998). A simple method for imposing local curvature in some flexible consumer-demand systems. Journal of Business and Economic Statistics, 16, 331--338.

Serletis A, Shahmoradi A. (2007). A note on imposing local curvature in generalized Leontief models. Macroeconomic Dynamics, 11, 290--294.

Serletis, A., Shahmoradi, A.(2007). Flexible functional forms, curvature conditions, and the demand for assets. Macroeconomic Dynamics, 11:455--486.

Shen, X., Yao, Z., Huo, H., Hea, K., Zhang, Y., Liu, H., Ye, Y.(2014). PM2.5 emissions from light-duty gasoline vehicles in Beijing, China. Science of the Total Environment, 487, 521--527.

Stone , J.R.N. (1954a).The Measurement of Consumers' Expenditure and Behaviour in the United Kingdom, 1920--1938. Cambridge University Press.

Stone , J.R.N. (1954b). Linear Expenditure Systems and Demand Analysis:An Application to the Pattern of British Demand. Economic Journal, 64, 511--527.

Strotz, R.H.(1957). The Empirical Implications of a Utility Tree. Econometrica, 25, 269--280.

Sun, C. W., Ouyang, X. L.(2016). Price and expenditure elasticities of residential energy demand during urbanization: An empirical analysis based on the household-level survey data in China. Energy Policy, 88, 56--63.

Theil H. (1965). The Information Approach to Demand Analysis.Econometrica, 33, 67--87.

Wadud, Z. (2016). Diesel demand in the road freight sector in the UK: Estimates for different vehicle types. Applied Energy, 165, 849--857.

Wang, Q.-d., He, K.-b., Huo, H., James, L.(2005). Real-world vehicle mission factors in Chinese metropolis city Beijing. Journal of Environmental Sciences, 17, 319--326.

Wolff, H., Heckelei, T.(2010). Imposing Curvature and Monotonicity on Flexible Functional Forms: An Efficient Regional Approach. Computational Economics, 36, 309--339.

Yang, S., He, L.Y.(2016). Fuel demand, road transport pollution emissions and residents' health losses in the transitional China. Transportation Research Part D: Transport and Environment, 42, 45--59.

Zhang, S., Wu, Y., Liu, H., Huang, R., Yang, L., Li, Z., Fu, L., Hao, J.(2014). Real-world fuel consumption and CO2 emissions of urban public buses in Beijing. Applied Energy, 113, 1645--1655.
}

\end{document}